\documentclass[twocolumn,prl,amsmath,amssymb,floatfix,reprint,square,comma,numbers]{revtex4-1}
\usepackage{graphicx}
\usepackage{epstopdf}
\usepackage{epsfig}
\usepackage{ulem}
\usepackage{amstext}
\usepackage{makeidx}
\usepackage[colorlinks=true,linkcolor=black]{hyperref}
\usepackage{color,soul}
\usepackage{amsmath}
\usepackage{braket}

\begin{document}

\title{Emulating many-body localization with a superconducting quantum processor}

\author{Kai Xu$^{1}$}
\thanks{K. X. and J.-J. C. contributed equally to this work.}
\author{Jin-Jun Chen$^{2,3}$}
\thanks{K. X. and J.-J. C. contributed equally to this work.}
\author{Yu Zeng$^{2,3}$}
\author{Yuran Zhang$^{2,3}$}
\author{Chao Song$^1$}
\author{Wuxin Liu$^1$}
\author{Qiujiang Guo$^1$}
\author{Pengfei Zhang$^1$}
\author{Da Xu$^1$}
\author{Hui Deng$^2$}
\author{Keqiang Huang$^{2,3}$}
\author{H. Wang$^{1,4}$}
\email{hhwang@zju.edu.cn}
\author{Xiaobo Zhu$^4$}
\email{xbzhu16@ustc.edu.cn}
\author{Dongning Zheng$^{2,3}$}
\author{Heng Fan$^{2,3}$}
\email{hfan@iphy.ac.cn}
\affiliation{ $^{1}$ Department of Physics, Zhejiang University, Hangzhou, Zhejiang 310027, China}
\affiliation{ $^{2}$ Beijing National Laboratory for Condensed Matter Physics,
Institute of Physics, Chinese Academy of Sciences, Beijing 100190, China}
\affiliation{ $^{3}$ School of Physical Sciences, University of Chinese Academy of Sciences, Beijing 100049, China}
\affiliation{ $^{4}$ Synergetic Innovation Centre in Quantum Information and Quantum
Physics, University of Science and Technology of China, Hefei, Anhui 230026, China}

\date{\today }

\begin{abstract}
The law of statistical physics dictates that generic closed quantum many-body systems
initialized in nonequilibrium will thermalize under their own dynamics.
However, the emergence of many-body localization (MBL) owing to the interplay
between interaction and disorder, which is in stark contrast to Anderson localization that only addresses
noninteracting particles in the presence of disorder,
greatly challenges this concept because it prevents the systems from evolving to the ergodic thermalized state.
One critical evidence of MBL is the long-time logarithmic growth of entanglement entropy,
and a direct observation of it is still elusive due to the experimental challenges
in multiqubit single-shot measurement and quantum state tomography.
Here we present an experiment of fully emulating the MBL dynamics
with a 10-qubit superconducting quantum processor, which represents a spin-1/2 XY model
featuring programmable disorder and long-range spin-spin interactions.
We provide essential signatures of MBL, such as the imbalance due to the initial nonequilibrium,
the violation of eigenstate thermalization hypothesis, and,
more importantly, the direct evidence of the long-time logarithmic growth of entanglement entropy.
Our results lay solid foundations for precisely
simulating the intriguing physics of quantum many-body systems
on the platform of large-scale multiqubit superconducting quantum processors.
\end{abstract}

\maketitle

{\it Introduction.}---
A central assumption of statistical mechanics is that generic closed
quantum systems driven out of equilibrium
will thermalize to the ergodic state which has no quantum correlations~\cite{PhysRevA.43.2046,PhysRevE.50.888,rigol2008thermalization,gogolin2016equilibration}.
One exception was demonstrated by Anderson~\cite{anderson1958absence}, who argued
that disordered systems featuring single-particle localization, known as Anderson localization, can fail to thermalize.
Systems exhibiting Anderson localization require noninteracting particles with low excitation energies,
and have been widely studied in a number of works~\cite{wiersma1997localization,dalichaouch1991microwave,
scheffold1999localization,chabanov2000statistical,schwartz2007transport,kondov2011three}.
However, in quantum many-body systems featuring interacting particles and high energy excitations
where Anderson localization is no longer applicable,
there emerges a new phase of localization, many-body localization (MBL)~\cite{basko2006metal},
which also prevents the systems from thermalizing and breaks down ergodicity.
The MBL phase resembles the Anderson localization phase in that both phases
explicitly go against the eigenstate thermalization hypothesis (ETH),
which implies that entanglement entropy 
violates volume law~\cite{kjall2014many,nandkishore2015many}.
Nevertheless, the MBL phase has very different dynamical properties~\cite{PhysRevLett.113.147204},
and a unique signature of MBL is the long-time logarithmic growth of entanglement entropy,
which correlates with a slow evolution towards equilibrium as resulting from dephasing
caused by interactions between particles~\cite{vznidarivc2008many,ponte2015many,modak2015many,levi2016robustness}.

Recent experimental progresses have allowed the realization of MBL in a controllable manner
on various artificially-engineered platforms, which have facilitated the detailed investigations
of thermalization and MBL in quantum many-body systems covering a wide range of aspects,
such as the emergence of the disorder-induced insulating state~\cite{kondov2015disorder},
the breaking down of ergodicity~\cite{schreiber2015observation}, and
the difference between Anderson localization and MBL in optical lattice \cite{bordia2016coupling}, and the
localization-delocalization transition in a three-dimensional system with nuclear magnetic resonance~\cite{alvarez2015localization}.
Moreover, the long-time logarithmic growth of entanglement entropy, the hallmark of MBL, is indirectly
shown by measuring the quantum Fisher information in a disordered spin chain with 10 trapped ions \cite{smith2016many}.
However, a direct observation of the MBL hallmark requires the capability of
performing fast and accurate quantum state tomography (QST) on the many-body system, which
has yet to be achieved.

\begin{figure*}[t]
\includegraphics[width=5.0in]{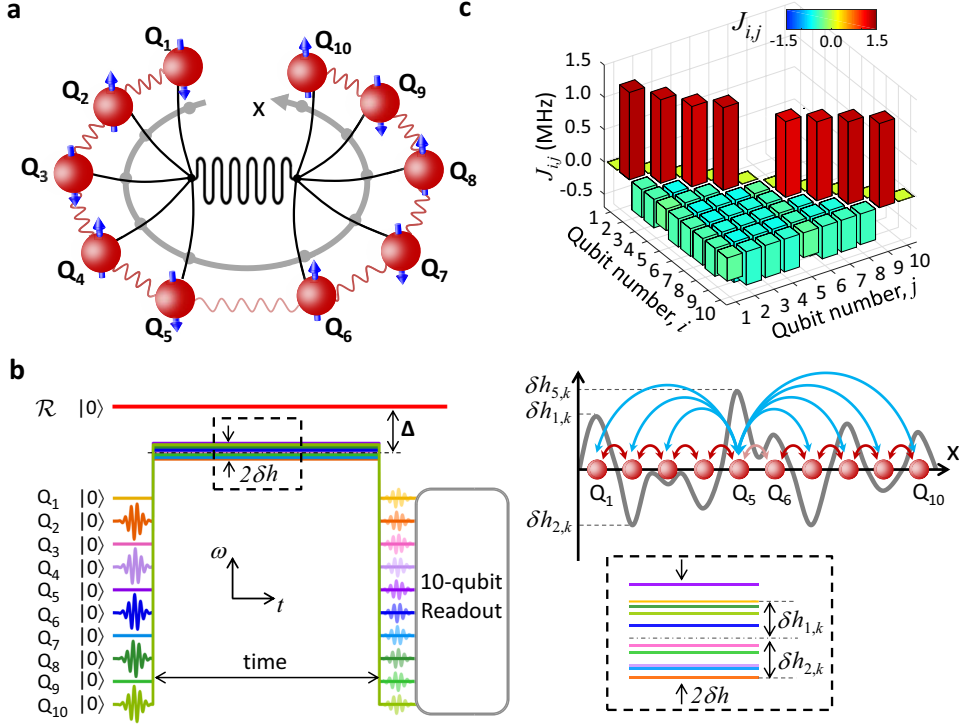}
\caption{\textbf{Experimental setup.}
\textbf{a,} Diagram of the 10-qubit superconducting quantum processor.
The qubits, shown as atoms with spins initialized in alternate orientations,
are arranged along a circular chain with the nearest-neighbor couplings represented
by red wavy lines, which are calibrated in a separate measurement as discussed in Supplementary Material
(the coupling for $Q_5$ and $Q_6$ is smaller than others).
The long-range, i.e., beyond nearest-neighbor, spin-spin interactions
are enabled by the central bus resonator $\mathcal{R}$
that couples to each individual qubits as illustrated by curved connecting lines.
\textbf{b,} Pulse sequences for emulating MBL. All 10 qubits are initialized at
their respective idle frequencies by applying $\pi$ pulses (dark-color sinusoids)
to the even-site qubits to prepare the initial state  $\vert 0101010101\rangle$,
following which the rectangular pulses are applied to quickly bias all qubits
to {\it nearby} $\Delta/2\pi = -650$~MHz. 
Individual Q$_{i}$ is offset from $\Delta$ by a small amount of $\delta h_{i,k} \in [-\delta h,\, \delta h]$,
where $\delta h_{i,k}$ is randomly chosen but fixed for the $k$-th pulse sequence
(see the bottom right panel for the zoomed-in view of the pulse segment enclosed by dashed lines), and
the ensemble of the $k = 1$ to 30 pulse sequences effectively emulates the random disordered potential $\delta h_{i}$. 
After the 10-qubit system evolves for a specific time from 0 to 1000~ns under the square pulses, 
all qubits are biased back to their respective idle frequencies for the 10-qubit joint readout,
which returns binary outcomes of the 10 qubits: We run all the $k =1$ to 30 sequences, each being executed for 3000 times,
to count $2^{10}$ probabilities of \{$P_{00\ldots0}$, $P_{00\ldots1}$, $\ldots$, $P_{11\ldots1}$\}$_k$ for $k = 1$ to 30.
Mean of the $k = 1$ to 30 probability ensemble captures the effect of the random disorder.
If the $N$-qubit QST is necessary, we insert tomographic rotation pulses (light-color sinusoids)
to the involved qubits before the $N$-qubit joint readout
to obtain all tomographic probabilities, $3^N$ more than aforesaid, to calculate the $N$-qubit density matrix for the $k$-th sequence.
Top right: The circular spin chain arranged in one dimension for illustrating the quantity $\delta h_{i,k}$.
\textbf{c,} The spin-spin coupling matrix $J_{ij}$ for $\Delta/2\pi = -650$~MHz.
Only nearest-neighbor couplings are positive,
while all other interaction terms are negative, which do not decay over distance.}
\label{fig:Hamiltonian}
\end{figure*}

Here we present an experiment of fully emulating the MBL dynamics
with a superconducting quantum processor, which represents a spin-1/2 XY model
featuring tunable disorder and long-range spin-spin interactions.
Our processor chip integrates 10 frequency-tunable transmon qubits that are interconnected by a central
bus resonator $\mathcal{R}$, with the circuit architecture introduced in Fig.~\ref{fig:Hamiltonian} and elsewhere~\cite{Song2017}.
Some prominent characteristics of this experimental platform for MBL are as follows.
First, the frequency and the state of each qubit can be individually manipulated via its own control lines,
and the qubit-qubit interaction between arbitrary two qubits can be mediated by detuning their frequency
from that of the resonator $\mathcal{R}$, so that both the disorder and the long-range interactions
are programmable. This is a necessary condition for MBL
since the XY model becomes non-integrable with the nonvanishing long-range interactions.
Second, the fast and accurate QST as demonstrated for up to 10 superconducting qubits in Ref.~\cite{Song2017}
allows us to record the dynamics of entanglement entropy, making it possible for
observing the aforesaid MBL hallmark.
Third, with the recent advances in coherence, scalability, and controllability
for superconducting quantum circuits~\cite{dicarlo2009demonstration,
mariantoni2011implementing,fedorov2012implementation,barends2014superconducting,Riste2015,Takita2016,Song2017,Rosenberg2017},
the platform becomes well suited for simulating and exploring MBL and
other intriguing but intractable questions of quantum many-body systems~\cite{Georgescu2014}.\\

{\it Hamiltonian.}---
In our superconducting quantum processor, the resonator mediated
super-exchange interactions between arbitrary two qubits give an
effective Hamiltonian as (see Fig.~\ref{fig:Hamiltonian}, Supplementary Material, and Refs.~\cite{Song2017}),
\begin{equation}
\frac{H}{\hbar} = \sum_{i<j}J_{ij}(\sigma_{i}^+\sigma_{j}^- + \sigma_{i}^-\sigma_{j}^+ )+\sum_{i}(h_i+\delta h_i)\sigma_{i}^+\sigma_{i}^-,
\label{eq:no1}
\end{equation}
where $\sigma_{i}^{\pm}$ are the raising/lowering operators, $h_i$ is the strength of the inherent transverse magnetic field,
and $\delta h_i$ is the random disordered potential of the $i$-th spin, and
$J_{ij}$ is the coupling strength between the $i$-th and $j$-th spins. The Hamiltonian is an effective XY model,
which conserves the total number of spin excitations 
\cite{jurcevic2014quasiparticle}.

As shown in Fig.~\ref{fig:Hamiltonian}, $J_{ij}$ contains two parts,
the nearest-neighbor direct coupling term $\lambda^\textrm{c}_{i,i+1}$, and the super-exchange interaction $J^\textrm{se}_{ij}$.
Most $\lambda^\textrm{c}_{i,i+1}/2\pi$ are around 1.8~MHz (see Supplementary Material), which
play a leading role compared with the corresponding super-exchange interactions.
The super-exchange interaction $J^\textrm{se}_{ij}$ between arbitrary two qubits Q$_i$ and Q$_j$
arises only if the two qubits are biased to the same detuning, i.e., $\Delta_{i}=\Delta_{j}$,
with a magnitude $\propto 1/\Delta_{i}$. In this experiment all qubits are detuned to $\Delta/2\pi \approx -650$ MHz,
therefore $J^\textrm{se}_{ij}/2\pi$ range from $-0.33$ MHz to $-0.64$ MHz.
The spin-spin coupling matrix $J_{ij}$ is shown in Fig.~\ref{fig:Hamiltonian}c.

The inherent magnetic field strength $h_i$ ($\propto 1/\Delta_{i}$) remains constant in the experiment~(see Supplementary Material and Ref.~\cite{Song2017}).
$\delta h_i$ denotes the on-site disorder potential taken from a uniform random distribution with
$\delta h_i \in [-\delta h,\,\delta h]$. This disorder is generated by applying the frequency shift of $\delta h_i$
to Q$_i$, for $i = 1$ to 10, on top of the large detuning $\Delta$,
where $\delta h \ll \Delta$ so that the coupling matrix $J_{ij}$ in Fig.~\ref{fig:Hamiltonian}c remains invariant.
Experimentally, evolution of the system towards either thermalization or the MBL phase is controlled
by the disorder strength $\delta h$ (see Fig.~\ref{fig:Hamiltonian}).\\

\begin{figure}[t]
\includegraphics[width=3.4in]{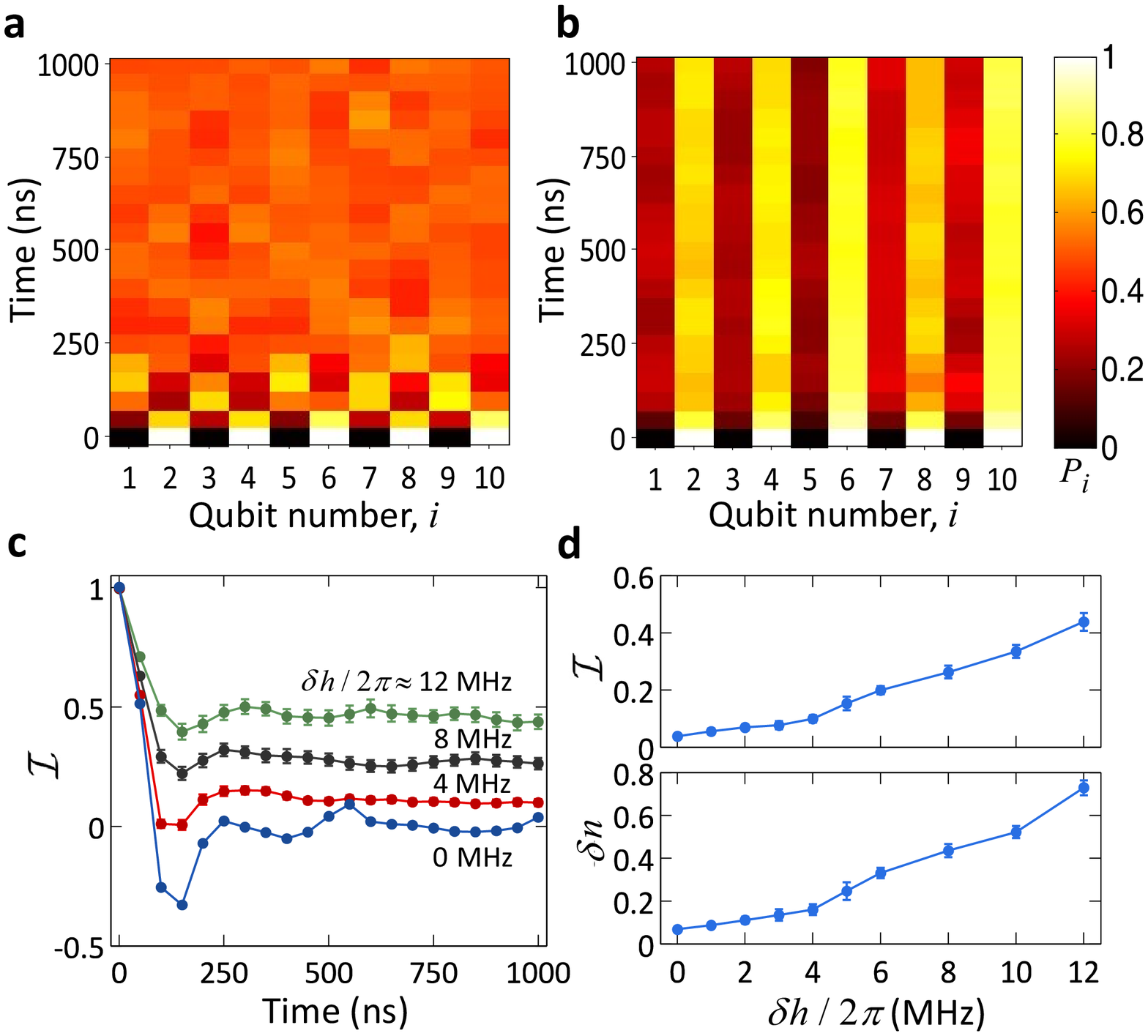}
\caption{\textbf{Dynamics of imbalance} for the system initialized in $\vert 0101010101\rangle$.
Recorded are the time evolutions of the excited-state ($\vert 1\rangle$-state) probabilities $P_i$ (colorbar on the far right)
for Q$_i$ (indexed as qubit number $i$) when there is no disorder with $\delta h\approx0$ (\textbf{a})
and a strong disorder with $\delta h/2\pi\approx12$ MHz (\textbf{b}). All probabilities are after readout correction~\cite{Zheng2017}.
\textbf{c,} Time evolutions of the system imbalance $\mathcal{I}$ at
different disorder strengths $\delta h$ as listed.
\textbf{d,}
The quasi-steady-state imbalance $\mathcal{I}$ (top) and
standard deviation $\delta n$ of the $|1\rangle$-state probability distributions
taken at 1000~ns as functions of $\delta h$ showing
that the system thermalizes for no disorder but starts to enter the MBL phase with increasing disorder strength.
Error bars are 1 SD calculated from all probability data of the $k = 1$ to 30 pulse sequences. }
\label{fig:Imbalance}
\end{figure}

{\it Imbalance.}---
Emergence of imbalance and ergodicity breaking are important signatures for the system
crossing from the thermalized phase to the MBL phase.
In the experiment, we initialize the system by preparing a 10-qubit N\'{e}el ordered state,
$\vert\psi_0\rangle=\vert 0101010101\rangle$,
with $\vert 0\rangle$ representing the ground state of a qubit on odd number sites
and $\vert 1\rangle$ representing the excited state on even number sites.
We study the ergodic properties via tracing the system imbalance due to the N\'{e}el nonequilibrium,
defined as $\mathcal{I}=\frac {N_\textrm{e}-N_\textrm{o}}{N_\textrm{e}+N_\textrm{o}}$,
where $N_\textrm{e}$ ($N_\textrm{o}$)
is the total number of excitation quanta on the even (odd) number sites.


\begin{figure*}[t]
\includegraphics[width=7.0in]{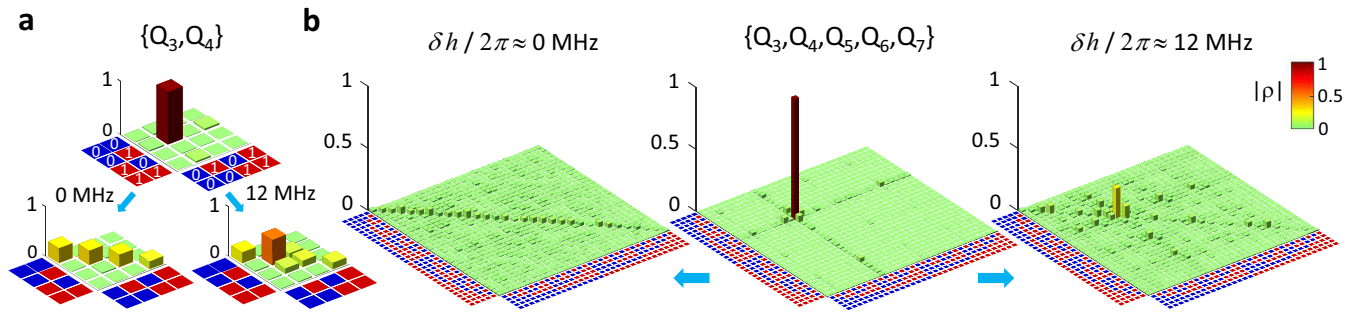}
\caption{\textbf{Subsystem density matrices for test of ETH.}
Shown are the experimental initial density matrices in absolute value for the two- (\textbf{a})
and five-qubit (\textbf{b}) subsystems in comparison with those
probed at 1000 ns when the system evolves to either the thermalized state ($\delta h \approx 0$) or the MBL phase ($\delta h/2\pi \approx 12$~MHz).
Density matrix data of the one-qubit subsystem Q$_3$ at a few selected evolution times
can be found in the Supplementary Material.
}
\label{fig:ETH}
\end{figure*}

We apply the pulse sequence as shown in Fig.~\ref{fig:Hamiltonian}b and record the system
dynamics with the 10-qubit joint readout. At long times around 250 ns or above, the system reaches a quasi-steady state.
Figures~\ref{fig:Imbalance}a and b show the time evolutions of the $\vert 1\rangle$-state probabilities of individual qubits,
which approach 0.5 for all qubit sites after $t\approx 250$ ns in the absence of disorder,
but maintain almost the original values in the presence of a strong disorder ($\delta h/2\pi\approx12$ MHz).
In Fig.~\ref{fig:Imbalance}c, it is seen that imbalance
$\mathcal{I}$ reaches a quasi-steady-state value at $\approx250$ ns for all disorder strengths.
The quasi-steady-state $\mathcal{I}$ is 0 for $\delta h\approx0$,
but remains non-zero and becomes larger if the disorder strength increases, signaling the entrance to the MBL phase that breaks down ergodicity.
The time needed in reaching the quasi-steady state, $\approx250$ ns, and the maximum evolution time we measured, $\approx1000$ ns,
are much less than the energy decay and dephasing times of our
superconducting qubits, both on the order of $10$ $\mu$s, which ensures that the effect of environment is almost negligible.
Meanwhile, we have post-selected the qubit probabilities that conserve the total excitations
to guarantee that our experimental system is an effectively closed quantum system (see Supplementary Material).

The quasi-steady-state imbalance can be taken as an order parameter to
quantify the crossover from the ergodic thermal phase to the non-ergodic MBL phase
tuned by the disorder strength $\delta h$, as shown in Fig.~\ref{fig:Imbalance}d.
We also measure the standard deviation $\delta n=\sqrt{\sum_{i=1}^{10}[P_i(0)-P_i(\delta h)]^2}$ of the $|1\rangle$-state
probability distributions between the ideal thermal state with $P_i(0)=0.5$, for $i = 1$ to 10, and the quasi-steady state
as a function of the disorder strength $\delta h$, which
works as a sensitive detector to witness the crossover.\\

{\it Eigenstate Thermalization Hypothesis.}---
Our initial N\'{e}el state is a product state, whose
arbitrary subsystem has only one nonzero element in the reduced density matrix.
In the absence of disorder, the N\'{e}el state evolves to the thermal state
which satisfies ETH and ensures that any resulting subsystem is a completely-mixed state
described by generalized Gibbs ensemble with an infinite temperature (see Supplementary Material).
In contrast, in the presence of a strong disorder, ETH fails and the reduced density matrix
of an arbitrary subsystem retains the initial form.

Using the multiqubit QST, we show in Fig.~\ref{fig:ETH} the averaged norms of the
subsystem density matrices at 0 and 1000~ns for two cases, one in the absence of disorder
with $\delta h\approx0$ and the other in the presence of a strong disorder with $\delta h/2\pi\approx12$ MHz.
For subsystems with one, two, and five qubits, the experimental data all agree reasonably well with with the expected thermal equilibrium
when $\delta h\approx0$, but retain the initial form when $\delta h/2\pi\approx12$ MHz, the latter of which clearly violates ETH.\\

{\it Entanglement entropy.}---
The dynamics of entanglement entropy for an isolated system
is a well defined signature for differentiating between
thermalization, Anderson localization, and MBL.
Here we focus on the evolution of the half-chain entanglement entropy using QST
for the system initialized in the N\'{e}el ordered state.
With inter-particle interactions but no disorder, the system is quickly thermalized and
its entanglement entropy saturates to the maximum (thermal) entropy
that depends on the system size and satisfies volume law~\cite{gogolin2016equilibration};
With strong disorder but no inter-particle interactions, there
arises Anderson location and entanglement entropy
quickly saturates to a constant that is independent of the system size
and much smaller than the maximum entropy~\cite{modak2015many};
With both strong disorder and inter-particle interactions,
the system enters the MBL phase, where the disorder prevents
the particle transport and leads to a slow growth of entanglement entropy
compared with that of the thermal phase,
but the inter-particle interactions contribute to the transport
of phase correlations so that entanglement entropy persistently
increases logarithmically in time compared with that of
Anderson localization~\cite{bardarson2012unbounded,pino2014entanglement,ponte2015many,modak2015many}.

It is argued that the slow logarithmic growth of entanglement entropy
is robust even for a quantum system under dissipation \cite{levi2016robustness}.
Without loss of generality, here we quote the 5 qubits \{Q$_3$, Q$_4$, Q$_5$, Q$_6$, Q$_7$\}
as subsystem $A$ and the rest of qubits as subsystem $B$,
and study the evolution of the half-chain entanglement
entropy $S=-{\rm tr}(\rho_A\ln\rho_A)$, where $\rho _A$ is the reduced density operator of subsystem $A$
by tracing out the subsystem $B$ (see Supplementary Material for a similar
set of experimental data with another choice of the subsystems).

\begin{figure}
\includegraphics[width=3.4in,clip=True]{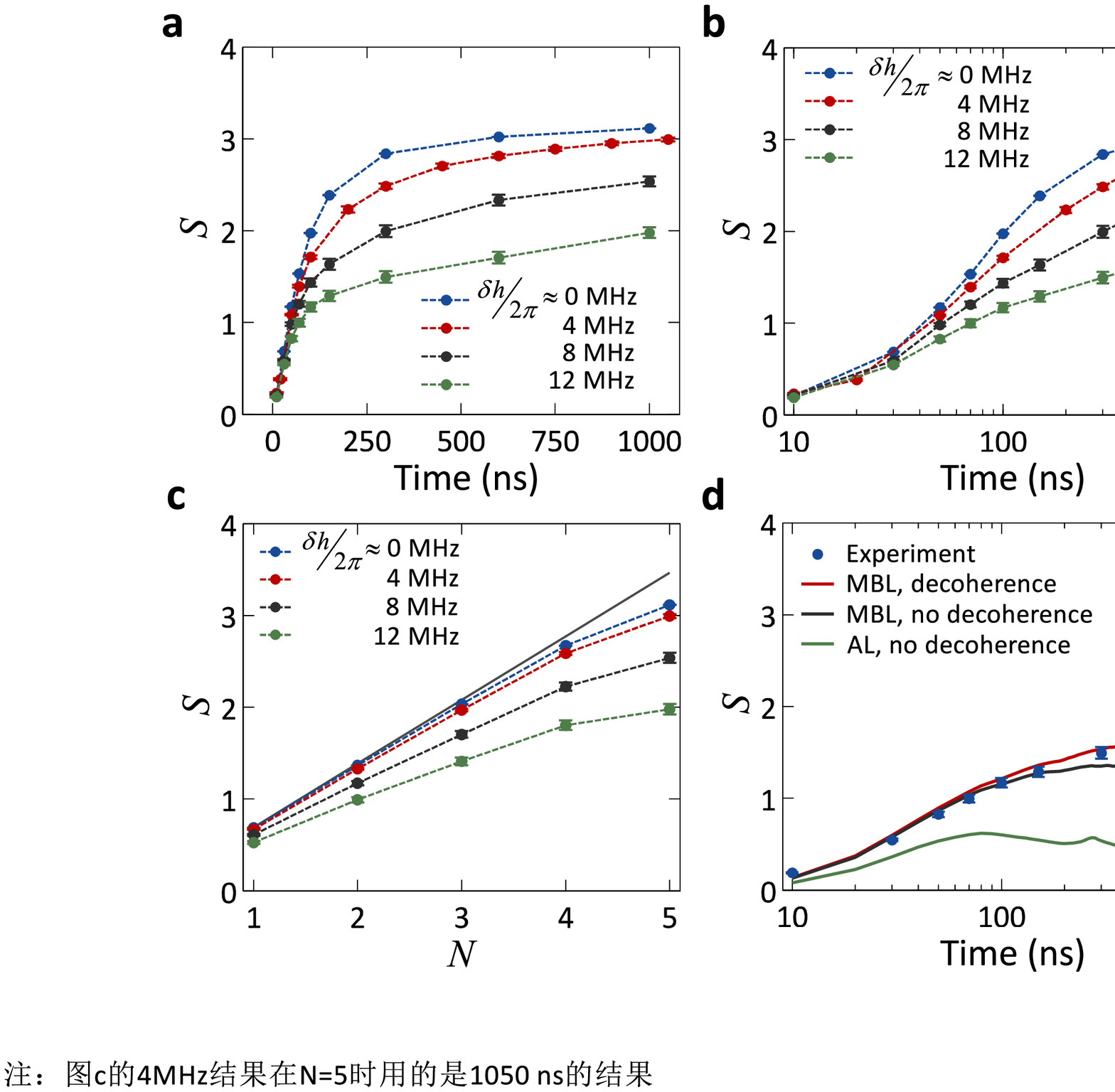}
\caption{\textbf{Half-chain entanglement entropy} for the 5-qubit subsystem
\{Q$_3$, Q$_4$, Q$_5$, Q$_6$, Q$_7$\}, of which the density matrices are obtained by QST.
Shown are the entanglement entropy $S$ as functions of
the evolution time, in linear scale (\textbf{a}) and in logarithmic scale (\textbf{b}),
at different disorder strengths $\delta h$ as labeled.
\textbf{c,} Site-averaged $S$ at around 1000~ns as functions the number of
sites (qubits) $N$ at different disorder strengths $\delta h$ as indicated,
which are calculated by taking the average of the entanglement entropies
of all $N$-site choices ($N\le5$) out of the directly measured 5-qubit subsystem.
Dots connected by dashed lines are experimental data. The solid line shows $S = N\ln 2$.
\textbf{d,} Entanglement entropy $S$ as functions of the
evolution time for comparison between MBL and Anderson localization (AL)
with the disorder strength $\delta h/2\pi\approx12$ MHz.
Dots are experimental data, and lines are numerical simulation results as indicated
(see Supplementary Material).
Error bars are $1$ SD calculated from all tomography data of the $k = 1$ to 30 pulse sequences.}
\label{fig:Entropy}
\end{figure}

Figures~\ref{fig:Entropy}a and b show the time evolutions
of the half-chain entanglement entropy at different disorder strengths as indicated.
As expected, for all disorder strengths, $S$ grows quickly and linearly in time at the beginning,
and then enters a slow growth period. For strong disorder strengths such as
$\delta h/2\pi \approx 8$ and 12~MHz, $S$ appears to grow logarithmically in time during the whole process.
Figure~\ref{fig:Entropy}c shows the site-averaged entanglement entropy at $\approx1000$~ns as functions
of the number of sites $N$ for different disorder strengths.
For $\delta h\approx0$ and $N \le 4$, the site-averaged entanglement entropy is close to thermal entropy, $N\ln {2}$,
which satisfies volume law. But for strong disorder such as $\delta h /2\pi \approx 12$~MHz,
it falls significantly below thermal entropy and therefore violates volume law.
Figure~\ref{fig:Entropy}d shows the difference between MBL and Anderson localization.
When the coupling strengths $J_{ij}$ contain only nearest-neighbor terms,
the Hamiltonian in Eq.~\ref{eq:no1} can be mapped to a
noninteracting fermionic model (see Supplementary Material),
and the strong disorder gives rise to Anderson localization, where
entanglement entropy saturates quickly (green line).
In contrast, the MBL phase demonstrates the long-time logarithmic growth of entanglement entropy,
as directly observed experimentally (dots), which is in excellent agreement with the numerical simulation
taking into account decoherence (red line).\\

In conclusion, we have presented key evidences for MBL and thermalization in an interacting many-body
system controllably induced by strong disorder and no disorder, respectively.
Our implementation is based on a 10-qubit superconducting quantum processor
which provides tunable disorder and arbitrary qubit-qubit interactions.
The interactions do not decay over distance and thus are nonlocal,
so that the MBL phase achieved in our setup can be viewed as supplemental
to the generally studied ``finite range'' situation~\cite{hastings2010locality}.
Furthermore, we have directly observed the long-time logarithmic growth of entanglement entropy,
the hallmark of the MBL phase. Our demonstration
shows that superconducting quantum processors can work precisely in simulating various intriguing
phenomena of quantum many-body systems.\\

This work was supported by the National Basic Research Program
of China (Grants No. 2014CB921201, No. 2016YFA0302104, and No. 2014CB921401), the
National Natural Science Foundations of China (Grants No. 11434008, No. 91536108, and No. 11374344),
and the Fundamental Research Funds for the Central Universities
of China (Grant No. 2016XZZX002-01).
Devices were made at the Nanofabrication
Facilities at Institute of Physics in Beijing, University
of Science and Technology of China in Hefei, and National
Center for Nanoscience and Technology in Beijing.

\clearpage

\newpage

\noindent {\bf Supplementary Material for: ``Emulating many-body localization with a superconducting quantum processor''}

\section{Hamiltonian}

The superconducting quantum processor used in this experiment is identical to the
one in reference \cite{song2017}, and more device information can be found
in its Supplementary Material. The system can be fully described by the Hamiltonian,
\begin{eqnarray}
\frac {H_1}{\hbar }&=& \omega_\mathcal{R} a^{\dagger}a + \sum_{i=1}^{10}\omega_{i}\sigma_{i}^+\sigma_{i}^-
+ \sum_{i=1}^{10}g_{i}(\sigma _{i}^{+}a+\sigma _{i}^{-}a^{\dagger}) \nonumber \\
&+&\sum_{i=1}^{9}\lambda^\textrm{c}_{i,i+1}(\sigma^{+}_{i}\sigma^{-}_{i+1}+\sigma^{-}_{i}\sigma^{+}_{i+1}),
\label{Seq:no1}
\end{eqnarray}
where $a^\dagger$ ($a$) is the creation (annihilation) operator of resonator
$\mathcal{R}$ and $\sigma_{i}^+$ ($\sigma_{i}^-$) is the raising (lowering) operator of the $i$-th qubit.
This Hamiltonian consists of 10 qubits and a resonator featuring the qubit-resonator couplings
$g_i$ and the nearest-neighbor qubit-qubit crosstalk couplings $\lambda^\textrm{c}_{i,i+1}$, the latter of which
are calibrated as listed in Table~\ref{tab1} using the qubit-qubit energy swaps at the many-body localization (MBL) operation point.

\begin{table}[b]
	\renewcommand\arraystretch{1.5}
	\begin{tabular}{c|ccccc}
		\hline
		\hline
		Qubit pair  &  Q$_1$-Q$_2$  &   Q$_2$-Q$_3$  &   Q$_3$-Q$_4$  &   Q$_4$-Q$_5$  &   Q$_5$-Q$_6$ \\
		$\lambda_{i,i+1}^\textrm{c}/2\pi$ (MHz) &  1.8 & 1.9 & 1.9 & 1.8 &  0.1 \\
		\hline
		Qubit pair  &  Q$_6$-Q$_7$ &  Q$_7$-Q$_8$ &  Q$_8$-Q$_9$ &   Q$_9$-Q$_{10}$ &  Q$_{10}$-Q$_1$ \\
		$\lambda_{i,i+1}^\textrm{c}/2\pi$ (MHz) &  1.8 &  1.8 &  1.9 &  1.8 & 0.0 \\
		\hline
		\hline
	\end{tabular}
	\caption{\label{table1} \textbf{Nearest-neighbor couplings, $\lambda_{i,i+1}^\textrm{c}$,} calibrated using the qubit-qubit energy swaps
	at the MBL operation point ($\Delta/2\pi\approx -650$~MHz), which show small but noticeable differences from the previous values
	measured at the GHZ operation point ($\Delta/2\pi\approx -140$~MHz) in reference \cite{song2017}.
	The differences are likely attributed to the different cool-downs which caused a slight change of
	the sample properties and also the weak dependence of $\lambda^c_{i,i+1}$ on the qubit frequencies.}
	\label{tab1}
\end{table}

In addition to the direct couplings between nearest-neighbor qubits $\lambda_{i,i+1}^\textrm{c}$,
another direct qubit-qubit coupling of Q$_i$ and Q$_j$ can be realized by the super-exchange
interaction with the qubits and the resonator interacting in the dispersive regime, i.e., $g_{i}\ll|\Delta|\equiv\left|\omega_{i}-\omega_{\mathcal{R}}\right|$,
where no energy exchange between the qubits and resonator occurs.
When all qubits are biased to the uniform detuning point $\Delta$ with the resonator $\mathcal{R}$
initialized in vacuum state, the effective Hamiltonian can be written as the following
while the resonator remains in vacuum throughout the process,
\begin{eqnarray}
\frac {H_2}{\hbar } &=& \sum_{i=1}^{9}\sum_{j=i+1}^{10}J^\textrm{se}_{ij}(\sigma_{i}^+\sigma_{j}^-
+ \sigma_{i}^-\sigma_{j}^+ )+\sum_{i=1}^{10}h_i\sigma_{i}^+\sigma_{i}^-
\nonumber \\
&+&\sum_{i=1}^{9}\lambda^\textrm{c}_{i,i+1}(\sigma^{+}_{i}\sigma^{-}_{i+1}+\sigma^{-}_{i}\sigma^{+}_{i+1}),
\label{Seq:no2}
\end{eqnarray}
where the first term represents the super-exchange interaction $J^\textrm{se}_{ij}=g_i g_j/\Delta$, and
the second term is the inherent magnetic field strength $h_i=g_i^2/\Delta$.
The super-exchange interaction $J^\textrm{se}_{ij}$ and the nearest-neighbor coupling $\lambda^\textrm{c}_{i,i+1}$
add up to the complete spin-spin coupling, defined as $J_{ij}$ in equation (1) of the main text. In our experiment,
we set the large detuning $\Delta/2\pi\approx-650$~MHz.
Consequently the nearest-neighbor couplings $J_{i,i+1}/2\pi$ are in the range from 1.22 to 1.35~MHz,
except that $J_{5,6}/2\pi\approx -0.37$~MHz because $\lambda^\textrm{c}_{5,6}$ is small;
The long-range couplings $J_{i,j}/2\pi$, where $j\neq i\pm1$, spread in the range from $-0.64$ to $-0.33$~MHz.
The complete spin-spin coupling matrix $J_{ij}$ is shown in Fig.~1c of the main text.
The inherent magnetic fields $h_i/2\pi$ range from $-0.65$ to $-0.31$~MHz.

We generate the site-specific programmable disorder by applying to
each qubit a relatively small site-dependent frequency deviation
$\delta h_i$ based on the large detuning $\Delta$, where $\delta h_i$
follow the uniform random distribution $\left[-\delta h, \delta h \right]$.
The effective Hamiltonian takes the form as shown by equation (1) of the main text,
where the long-range couplings and the programmable disorder
are the two key ingredients to realize MBL.
In the experiment, we set different disorder strengths so that $\delta h/2\pi$ goes from 0 to 12~MHz,
and the variations of $J^\textrm{se}_{ij}$ and $h_i$
caused by different disorder strengths are negligible.

\begin{table*}[!htb]
	\centering
	\renewcommand\arraystretch{1.5}
	\begin{tabular}{c|cccccccccc}
		\hline
		\hline
		Qubits & Q$_1$ & Q$_2$ & Q$_3$ & Q$_4$ & Q$_5$ & Q$_6$ & Q$_7$ & Q$_8$ & Q$_9$ & Q$_{10}$ \\
		\hline
		$T_{1,i}$ ($\mu s$) & 25.6 & 21.6 & 9.8 & 14.3 & 14.2 & 32.5 & 11.9 & 9.4 & 17.9 & 30.6 \\
		\hline
		$T_{2,i}^{*}$ ($\mu s$) & 2.2 & 2.8 & 1.5 & 1.7 &  2.7 & 2.7 & 2.2 & 1.4 & 2.4 &  2.6 \\
		\hline
		$g_i/2\pi$ (MHz) & 14.2 & 20.5 & 19.9 & 20.2 & 15.2 & 19.9 & 19.6 & 18.9 & 19.8 & 16.3 \\
		\hline
		$\omega_{i,\textrm{idle}}/2\pi$ (GHz) & 5.114 & 5.459 & 5.600 & 5.043 & 5.648 & 5.166 & 4.958 & 5.248 & 5.081 & 5.554 \\
		\hline
		\hline
	\end{tabular}
	\caption{\label{table2} \textbf{Qubits characteristics.} $T_{1,i}$ and $T_{2,i}^{\ast}$
	are the energy lifetime and the Ramsey (Gaussian) dephasing time, respectively, of the $i$-th qubit (Q$_i$)
	measured at the MBL operation point. $\omega_{i,\textrm{idle}}$ is the idle frequency
	where the state preparation and measurement are executed. $g_i/2\pi$ is the coupling strength between
	Q$_i$ and resonator $\mathcal{R}$ \cite{song2017}.}
\end{table*}

\section{Effect of decoherence}

Although MBL assumes an isolated quantum system, which is hardly attainable in realistic conditions,
it is still possible to observe the MBL effect in a system that demonstrates excellent coherence performance
within the time window of the MBL dynamics. We conclude that our system can be approximately
regarded as a closed quantum system within the experimental time of around 1~$\mu$s, since the coupling rate
to environment is much slower than the localization rate of the system.
The coupling of the system to environment takes two forms, energy relaxation and dephasing,
and the characteristics for each individual qubits are shown in Table~\ref{table2}.
The energy lifetime $T_{1,i}$ of each qubit is indeed much longer than 1~$\mu$s.
The Ramsey (Gaussian) dephasing time $T_{2,i}^{\ast}$ measured for each individual qubit,
while all other qubits are far detuned, is comparable to the experimental time.
However, we believe that the effective dephasing time for each qubit that describes
the MBL dynamics shall be much longer, since during the MBL dynamics all qubits are coupled
and form a new system with eigenenergies that depend very weakly on each qubit flux.

\begin{figure}[t]
	\centering
	\includegraphics[width=2.5in,clip=True]{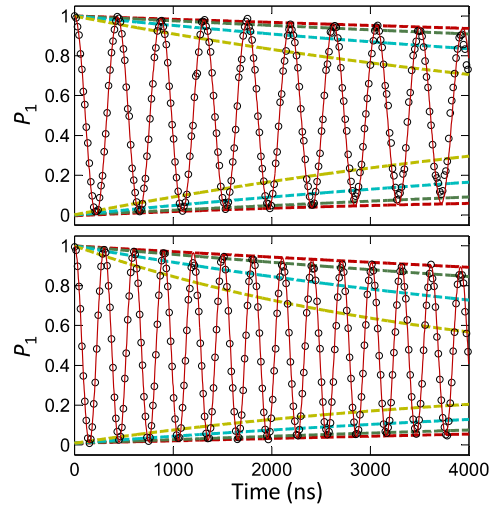}
	\caption{\textbf{Characterization of the pure dephasing time $T_{\varphi,i}$ for coupled-qubit systems.}
	\textbf{a,}~$|1\rangle$-state probability of Q$_7$ (circles), after readout correction, as a function
	of the interaction time during the energy swap between Q$_7$ and Q$_8$ at the MBL operation point.
	Q$_7$ is initialized in $|1\rangle$, following which both qubits are tuned to
	the MBL operation point and the excitation energy swaps between the two qubits.
	The solid red line is a numerical simulation based on
	the Lindblad master equation using the experimental $T_{1,i}$ values
	and $T_{\varphi,i} = 30\,\mu$s for $i=7$ and 8.
	Dashed lines are numerical simulations outlining the envelopes of the probability oscillations
	by parametrizing $T_{\varphi,i}$ with different colors:
	$T_{\varphi,i}$ for $i = 7$ and 8 are selected as 30 $\mu $s (red), 20 $\mu $s (green), 10 $\mu $s (cyan), and 5 $\mu $s (yellow).
  It is seen that the oscillation envelopes of $T_{\varphi,i}=30\,\mu$s agree with the experimental data the best.
	\textbf{b,}~$|1\rangle$-state probability of Q$_8$ (circles), after readout correction, as a function
	of the interaction time during the energy swap between Q$_7$, Q$_8$, and Q$_9$ at the MBL operation point.
	Compared with \textbf{a}, the oscillation frequency increases as expected.
	Lines are numerical simulations as described in \textbf{a}.
	It is seen that the oscillation envelopes of $T_{\varphi,i}=30\,\mu$s agree with the experimental data the best.
  \label{Exfig:T2swap}}
\end{figure}

We numerically model the MBL dynamics using the Lindblad master equation with the $T_{1,i}$
values given in Table~\ref{table2}, while the pure dephasing time $T_{\varphi,i}$ required
for each qubit in the simulation is set to 30~$\mu$s in order to explain the observed dynamics.
To justify our choice of long $T_{\varphi,i}$, we track the energy swap dynamics involving two and three qubits,
where the first qubit is excited to $|1\rangle$ and the rest group of qubits in $|0\rangle$ is
tuned to the same detuning for an energy exchange process. The $|1\rangle$-state probabilities
of the first qubit as functions of the swap time are recorded for both cases,
and compared with numerical simulations. Indeed we find that $T_{\varphi,i} \approx 30\,\mu$s ensures
a good agreement between the experiment and the simulation (Fig.~\ref{Exfig:T2swap}).

\begin{figure}[t]
	\centering
	\includegraphics[width=3.4in,clip=True]{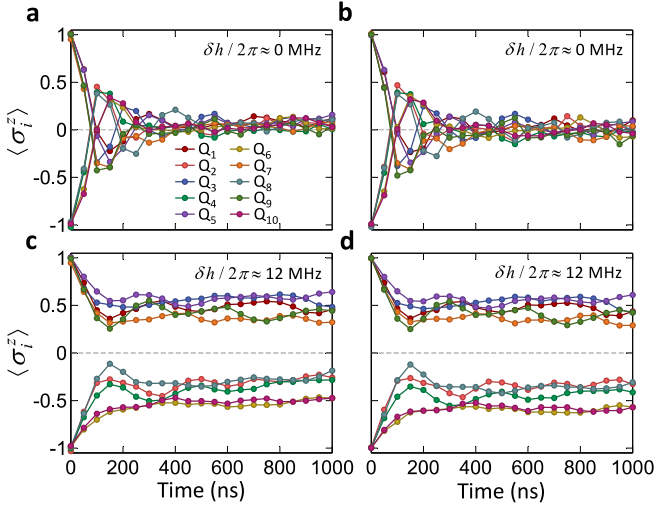}
	\caption{\footnotesize{\textbf{Time evolutions of the single-site magnetizations} with (right)
	and without (left) the post-selection for conservation of the total excitations
	at different disorder strengths $\delta h / 2\pi \approx 0$ (\textbf{a} and \textbf{b})
	and 12 MHz (\textbf{c} and \textbf{d}).
	The difference between the left and right panels are negligible, indicating
	that energy relaxation plays little effect during the 1~$\mu$s dynamics.}
  \label{Exfig:singleMag}}
\end{figure}

Furthermore, since the Hamiltonian conserves $\sum_{i}{\sigma_i^z}$ during the evolution,
we can post-select the states that conserve the total excitations to wipe out the effect of energy relaxation.
The effect of energy relaxation can be intuitively observed
by comparing the results of the single-site magnetization measurement
with and without post-selection. As shown in Fig.~\ref{Exfig:singleMag},
the comparison indicates that energy relaxation causes each qubit to decay from $|1\rangle$
to $|0\rangle$ and the expectation value of $\langle \sigma_i^z \rangle$ to drift from
-1 to 1. But such a drift is almost negligible during the 1~$\mu$s experimental time
due to the small energy relaxation rate of each qubit.

\section{Different initial state}

\begin{figure}[t]
	\begin{center}
		\includegraphics[width=3.4in,clip=True]{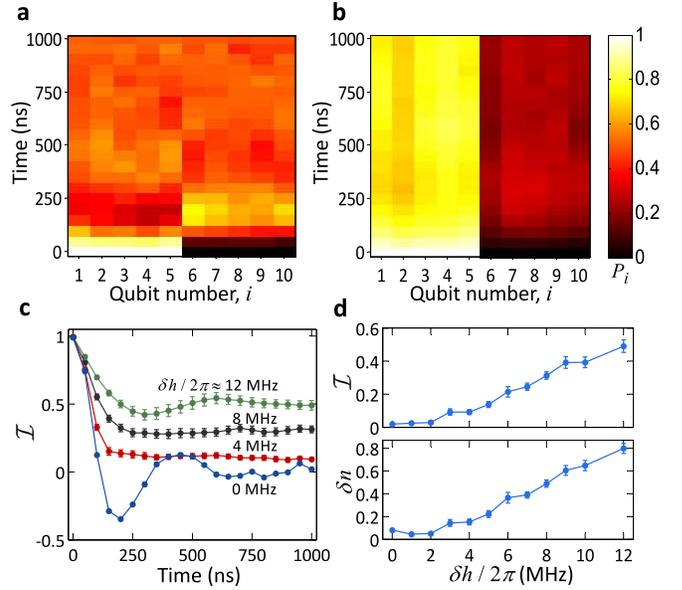}
		\caption{\textbf{Dynamics of imbalance}
			for the system initialized in $\vert\psi_0\rangle=\vert 1111100000\rangle$.
			Recorded are the time evolutions of the excited-state ($\vert 1\rangle$-state) probabilities $P_i$ (colorbar on the far right)
			for Q$_i$ (indexed as qubit number $i$) when there is no disorder with $\delta h\approx0$ (\textbf{a})
			and a strong disorder with $\delta h/2\pi\approx12$ MHz (\textbf{b}). All probabilities are after readout correction.
			\textbf{c,} Time evolutions of the system imbalance $\mathcal{I}$ at different disorder strengths $\delta h$ as listed.
			\textbf{d,}
			The steady-state imbalance $\mathcal{I}$ (top) and
			standard deviation $\delta n$ of the $|1\rangle$-state probability distributions
			taken at 1000~ns as functions of $\delta h$ showing
			that the system thermalizes for no disorder but starts to enter the MBL phase with increasing disorder strength.
			}
		\label{Exfig:Imbalance}
	\end{center}
\end{figure}

In order to show the effectiveness of imbalance for quantifying MBL,
we initialize the system in a different product state $\vert\psi_0\rangle=\vert 1111100000\rangle$,
which has a sharp density domain wall as shown in Fig.~\ref{Exfig:Imbalance}a.
The imbalance takes the form $\mathcal{I}=\left(N_\textrm{L}-N_\textrm{R}\right)/\left(N_\textrm{L}+N_\textrm{R}\right)$,
where $N_\textrm{L}$ is the number of excitations for the 5 qubits on the left side (Q$_1$ to Q$_5$), and
$N_\textrm{R}$ is the number of excitations for the 5 qubits on the right side (Q$_6$ to Q$_{10}$).
We introduce the same levels of disorders as those in the main text,
and observe similar dynamics towards either thermalization or MBL depending on the disorder strength.

Figure~\ref{Exfig:Imbalance}a,b shows the time evolution of the $|1\rangle$-state probabilities of all ten qubits.
In the absence of disorder, the system reaches a quasi-steady state and the domain wall disappears at around 350~ns (Fig.~\ref{Exfig:Imbalance}a).
The time needed to reach thermalization is slightly longer than the case presented in the main text
because of Lieb-Robinson bound \cite{lieb1972finite,jurcevic2014quasiparticle}.
In the presence of strong disorder with $\delta h/2\pi \approx 12$ MHz,
all excitations remain localized and the domain wall is clearly visible up to 1000~ns (Fig.~\ref{Exfig:Imbalance}b),
signifying the entrance to the MBL phase.

Figure~\ref{Exfig:Imbalance}c shows the time evolutions of imbalance at different disorder strengths.
The quasi-steady-state imbalance $\mathcal{I}$ is $0$ for $\delta h \approx 0$, but becomes larger as the disorder strength increases.
At 1000~ns, both $\mathcal{I}$ and the standard deviation $\delta n$ of the $|1\rangle$-state probability
distributions between the ideal thermal state and the quasi-steady state as functions of the disorder strength exhibit the crossover
from the ergodic thermal phase to the non-ergodic MBL phase (Fig.~\ref{Exfig:Imbalance}d).

\section{Eigenstate thermalization hypothesis}

For the generic closed quantum system, we divide the whole system into two subsystems, the relatively small A and the relatively large B.
In the absence of disorder, subsystem A would thermalize using subsystem B as a heat bath, and the thermalized
system satisfies eigenstate thermalization hypothesis (ETH)~\cite{PhysRevA.43.2046,PhysRevLett.80.1373,rigol2008thermalization}.
For a given Hamiltonian $H$ and the initial state $|\psi_0\rangle$, the energy $\langle\psi_0|H|\psi_0\rangle$ is equal to the thermal equilibrium energy,
\begin{eqnarray}
E=\frac{\textrm{tr}\left[He^{-\beta H}\right]}{\textrm{tr}\left[e^{-\beta H}\right]}, \quad \beta=\frac{1}{k_BT}.
\label{Seq:no5}
\end{eqnarray}
The reduced density matrix of subsystem A in thermal equilibrium can be estimated by
\begin{eqnarray}
\rho_\textrm{A}\equiv\rho_\textrm{A}(T)=\frac{\textrm{tr}_\textrm{B}\left[e^{-\beta H}\right]}{\textrm{tr}\left[e^{-\beta H}\right]}.
\label{Seq:no6}
\end{eqnarray}

In our experiment, for the Hamiltonian in equation~(1) of the main text and the N\'{e}el ordered initial state,
the energy $E=\langle\psi_0|H|\psi_0\rangle$ is a constant and $\beta=0$, i.e., $T \rightarrow \infty$, 
which means that the energy of the N\'{e}el ordered initial state is equivalent to a thermal state at an infinite temperature.
By equation~(\ref{Seq:no6}), the reduced density matrices for one qubit, e.g., $Q_3$, and two qubits, e.g., $\{Q_3, Q_4\}$,
in thermal equilibrium are, respectively,
\begin{equation}
\rho_1=
\left(
\begin{array}{cc}
1/2 & 0 \\
0 &1/2\\
\end{array}
\right),
\quad
\rho_2=
\left(
\begin{array}{cccc}
1/4& 0 & 0 & 0\\
0 & 1/4 & 0 & 0\\
0 & 0 & 1/4 & 0\\
0 & 0 & 0 & 1/4\\
\end{array}
\right).
\label{Seq:no7}
\end{equation}
The reduced density matrices are equivalent to the completely mixed states when its size is much smaller than the remainder of the system.

\begin{figure}[t]
	\centering
	\includegraphics[width=3.4in,clip=True]{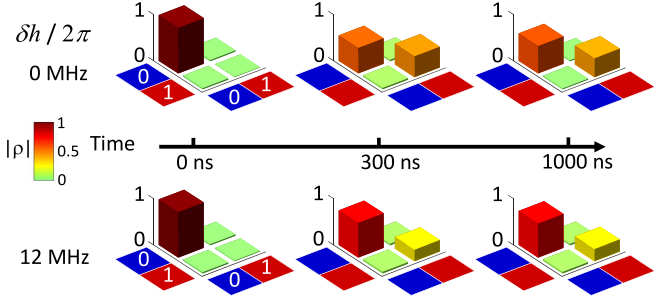}
	\caption{\textbf{Evolutions of the one-qubit subsystem}. The averaged
	density matrices in absolute value for Q$_3$ at the evolution times of 0, 300, and 1000~ns are
	shown for the two cases with and without disorder. The matrix difference between
	thermalization and MBL becomes significant starting from around 300~ns.
	\label{Exfig:ETH1qubit}}
\end{figure}

The density matrices of subsystem A at different evolutions times
under different disorder strengths are measured using quantum state tomography (QST).
When $\delta h\approx 0$, the experimental density matrices conform with the expectation of thermal equilibrium, including that for the subsystem A with five qubits;
when $\delta h/2\pi\approx 12$~MHz, the experimental density matrices
retain the initial form even at 1000~ns, as shown in Fig.~3 of the main text
and Fig.~\ref{Exfig:ETH1qubit}.

\section{Half-chain entanglement entropy}

The long-time logarithmic growth of the von Neumann entropy
is a distinctive feature of MBL. In this experiment, we achieve the direct measurement of
the half-chain von Neumann entropy with the 5-qubit QST \cite{song2017}.
The $5$-qubit density matrices of \{Q$_{3}$,Q$_{4}$,Q$_{5}$,Q$_{6}$,Q$_{7}$\} can be found in Fig.~3b of the main text.

\begin{figure}[t]
	\centering
	\includegraphics[width=3.4in,clip=True]{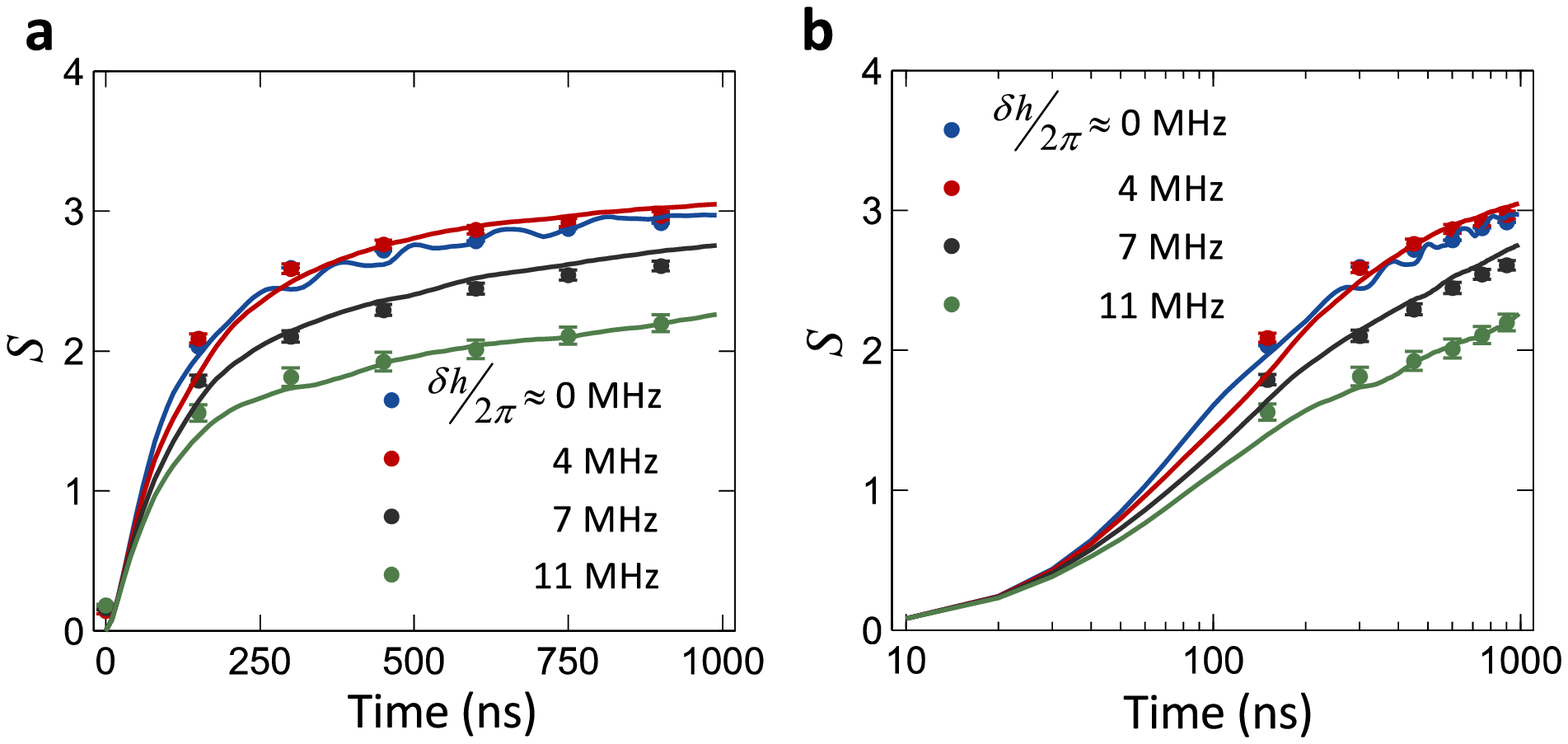}
	\caption{\textbf{Half-chain entanglement entropy} for the subsystem \{Q$_{1}$,Q$_{2}$,Q$_{3}$,Q$_{4}$,Q$_{5}$\}.
	Experimental data (dots) are plotted with the time axis in linear scale (\textbf{a}) and in logarithmic scale (\textbf{b}). Lines
	are numerical simulations with decoherence included.
	\label{Exfig:entropyq1Toq5}}
\end{figure}

To verify the robustness of this feature for entanglement entropy $S$,
we also perform the half-chain entropy measurement with the subsystem \{Q$_{1}$,Q$_{2}$,Q$_{3}$,Q$_{4}$,Q$_{5}$\},
and observe similar dynamics as shown in Fig.~\ref{Exfig:entropyq1Toq5}.
The half-chain entropy $S$ grows quickly at first and then enters the quasi-steady region
where $S$ increases slowly. Under strong disorders of $\delta h/2\pi\approx$ 7 and 11~MHz,
the logarithmic growth of $S$ in time is clearly visible, which agrees with numerical simulations taking into account decoherence.

\section{Effect of interactions}

The transport properties of the quantum system are determined
by the interplay between the disorder and the inter-particle interactions.
In the absence of interactions, the relevant physical model
is integrable and strong disorder gives rise to Anderson localization (single-particle localization), in which all single particle states are localized.
When considering the interactions of particles, the model is non-integrable,
which is the necessary condition for MBL.
For our effective Hamiltonian shown in equation~(1) of the main text, we can use the Jordan-Wigner transformation \cite{lieb1961two} to map the spin model to a fermionic model,
\begin{eqnarray}
\sigma_{i}^-&=&\exp\big[-i\pi\sum_{j=1}^{i-1}c_j^\dagger c_j\big]c_i \nonumber \\
\sigma_{i}^+&=&c_i^\dagger\exp\big[i\pi\sum_{j=1}^{i-1}c_j^\dagger c_j\big],
\label{Seq:no8}
\end{eqnarray}
where $c_i$ and $c_i^\dagger$ are fermionic operators, satisfying the anticommutation relation of
$\{c_i,c_j^\dagger\}=\delta_{ij}$ and $\{c_i,c_j\}=\{c_i^\dagger,c_j^\dagger\}=0$.
Because $c_j^\dagger c_j$ is an occupation number operator taking the eigenvalues $0$ and $1$,
$\exp(-i\pi c_j^\dagger c_j)=\exp(i\pi c_j^\dagger c_j)$.
The corresponding fermionic Hamiltonian is written as,
\begin{eqnarray}
\frac {H_3}{\hbar }&=&\sum_{i<j}J_{ij}\left(c_i^\dagger \exp\big[-i\pi\sum_{k=i}^{j-1}c_k^\dagger c_k\big]c_j+\textrm{h.c.}\right)\nonumber \\
&+&\sum_{i}\left(h_i+\delta h_i\right)c_i^\dagger c_i.
\label{Seq:no9}
\end{eqnarray}
We find that the Hamiltonian with long-range spin-spin couplings is an interacting fermionic model.
If $J_{ij}$ contains only nearest-neighbor spin-spin couplings, the Hamiltonian reduces into a noninteracting fermionic model.
Accordingly, we can numerically simulate Anderson localization using the experimental Hamiltonian
but setting all long-range spin-spin couplings to be zero,
which is plotted in Fig.~4d of the main text in support of the conclusion that
the long-time logarithmic growth of entanglement entropy in the MBL phase is indeed caused by inter-particle interactions.



\end{document}